\documentclass[twocolumn,aps,unsortedaddress,superscriptaddress,prl,floatfix,showpacs]{revtex4}
\usepackage{epsfig}

\begin{document}


\title{Measurement of Angular Distributions of Drell-Yan Dimuons
in $p + d$ Interaction at 800 GeV/c}

\affiliation{Abilene Christian University, Abilene, TX 79699}
\affiliation{Physics Division, Argonne National Laboratory, Argonne, IL 60439}
\affiliation{Fermi National Accelerator Laboratory, Batavia, IL 60510}
\affiliation{Georgia State University, Atlanta, GA 30303}
\affiliation{Illinois Institute of Technology, Chicago, IL  60616}
\affiliation{University of Illinois at Urbana-Champaign, Urbana, IL 61801}
\affiliation{Los Alamos National Laboratory, Los Alamos, NM 87545}
\affiliation{University of New Mexico, Albuquerque, NM 87131}
\affiliation{New Mexico State University, Las Cruces, NM 88003}
\affiliation{Oak Ridge National Laboratory, Oak Ridge, TN 37831}
\affiliation{Texas A\&M University, College Station, TX 77843}
\affiliation{Valparaiso University, Valparaiso, IN 46383}

\author{L.Y.~Zhu}
\affiliation{University of Illinois at Urbana-Champaign, Urbana, IL 61801}

\author{J.C.~Peng}
\affiliation{University of Illinois at Urbana-Champaign, Urbana, IL 61801}
\affiliation{Los Alamos National Laboratory, Los Alamos, NM 87545}

\author{P.E.~Reimer}
\affiliation{Physics Division, Argonne National Laboratory, Argonne, IL 60439}
\affiliation{Los Alamos National Laboratory, Los Alamos, NM 87545}

\author{T.C.~Awes}
\affiliation{Oak Ridge National Laboratory, Oak Ridge, TN 37831}

\author{M.L.~Brooks}
\affiliation{Los Alamos National Laboratory, Los Alamos, NM 87545}

\author{C.N.~Brown}
\affiliation{Fermi National Accelerator Laboratory, Batavia, IL 60510}

\author{J.D.~Bush}
\affiliation{Abilene Christian University, Abilene, TX 79699}

\author{T.A.~Carey}
\affiliation{Los Alamos National Laboratory, Los Alamos, NM 87545}

\author{T.H.~Chang}
\affiliation{New Mexico State University, Las Cruces, NM 88003}

\author{W.E.~Cooper}
\affiliation{Fermi National Accelerator Laboratory, Batavia, IL 60510}

\author{C.A.~Gagliardi}
\affiliation{Texas A\&M University, College Station, TX 77843}

\author{G.T.~Garvey}
\affiliation{Los Alamos National Laboratory, Los Alamos, NM 87545}

\author{D.F.~Geesaman}
\affiliation{Physics Division, Argonne National Laboratory, Argonne, IL 60439}

\author{E.A.~Hawker}
\affiliation{Texas A\&M University, College Station, TX 77843}

\author{X.C.~He}
\affiliation{Georgia State University, Atlanta, GA 30303}

\author{L.D.~Isenhower}
\affiliation{Abilene Christian University, Abilene, TX 79699}

\author{D.M.~Kaplan}
\affiliation{Illinois Institute of Technology, Chicago, IL  60616}

\author{S.B.~Kaufman}
\affiliation{Physics Division, Argonne National Laboratory, Argonne, IL 60439}

\author{S.A.~Klinksiek}
\affiliation{University of New Mexico, Albuquerque, NM 87131}

\author{D.D.~Koetke}
\affiliation{Valparaiso University, Valparaiso, IN 46383}

\author{D.M.~Lee}
\affiliation{Los Alamos National Laboratory, Los Alamos, NM 87545}

\author{W.M.~Lee}
\affiliation{Fermi National Accelerator Laboratory, Batavia, IL 60510}
\affiliation{Georgia State University, Atlanta, GA 30303}

\author{M.J.~Leitch}
\affiliation{Los Alamos National Laboratory, Los Alamos, NM 87545}

\author{N.~Makins}
\affiliation{Physics Division, Argonne National Laboratory, Argonne, IL 60439}
\affiliation{University of Illinois at Urbana-Champaign, Urbana, IL 61801}

\author{P.L.~McGaughey}
\affiliation{Los Alamos National Laboratory, Los Alamos, NM 87545}

\author{J.M.~Moss}
\affiliation{Los Alamos National Laboratory, Los Alamos, NM 87545}

\author{B.A.~Mueller}
\affiliation{Physics Division, Argonne National Laboratory, Argonne, IL 60439}

\author{P.M.~Nord}
\affiliation{Valparaiso University, Valparaiso, IN 46383}

\author{V.~Papavassiliou}
\affiliation{New Mexico State University, Las Cruces, NM 88003}

\author{B.K.~Park}
\affiliation{Los Alamos National Laboratory, Los Alamos, NM 87545}

\author{G.~Petitt}
\affiliation{Georgia State University, Atlanta, GA 30303}

\author{M.E.~Sadler}
\affiliation{Abilene Christian University, Abilene, TX 79699}

\author{W.E.~Sondheim}
\affiliation{Los Alamos National Laboratory, Los Alamos, NM 87545}

\author{P.W.~Stankus}
\affiliation{Oak Ridge National Laboratory, Oak Ridge, TN 37831}

\author{T.N.~Thompson}
\affiliation{Los Alamos National Laboratory, Los Alamos, NM 87545}

\author{R.S.~Towell}
\affiliation{Abilene Christian University, Abilene, TX 79699}

\author{R.E.~Tribble}
\affiliation{Texas A\&M University, College Station, TX 77843}

\author{M.A.~Vasiliev}
\affiliation{Texas A\&M University, College Station, TX 77843}

\author{J.C.~Webb}
\affiliation{New Mexico State University, Las Cruces, NM 88003}

\author{J.L.~Willis}
\affiliation{Abilene Christian University, Abilene, TX 79699}

\author{D.K.~Wise}
\affiliation{Abilene Christian University, Abilene, TX 79699}

\author{G.R.~Young}
\affiliation{Oak Ridge National Laboratory, Oak Ridge, TN 37831}

\collaboration{FNAL E866/NuSea Collaboration}
\noaffiliation

\date{\today}

\begin{abstract}
We report a measurement of the angular distributions of Drell-Yan
dimuons produced using an 800 GeV/c proton beam on a deuterium 
target. The muon angular distributions in polar angle $\theta$ and
azimuthal angle $\phi$ have been measured over the kinematic range $4.5 < 
m_{\mu \mu} < 15$ GeV/c$^2$,
$0 < p_T < 4 $ GeV/c, and $0 < x_F < 0.8$. No significant cos$2\phi$ dependence
is found in these proton-induced Drell-Yan data, in contrast to 
the situation for pion-induced Drell-Yan. The data are compared with
expectations from models which attribute 
the cos$2\phi$ distribution 
to a QCD vacuum effect or to the
presence of the transverse-momentum-dependent Boer-Mulders 
structure function $h_1^\perp$.
Constraints on the magnitude of the sea-quark 
$h_1^\perp$ structure functions are obtained.

\end{abstract} 

\maketitle
The Drell-Yan process~\cite{drell70}, in which a charged 
lepton pair is produced in a high-energy
hadron-hadron interaction via the $q \bar q \to l^+ l^-$ process, has been
a testing ground for perturbative QCD and a unique tool for probing parton
distributions of hadrons. The Drell-Yan production cross sections can be
well described by next-to-leading order QCD calculations~\cite{stirling93}. 
This provides 
a firm theoretical framework for using the Drell-Yan process to determine
the antiquark content of nucleons and nuclei~\cite{pat99}, 
as well as the quark
distributions of pions, kaons, and antiprotons~\cite{kenyon82}. 

Despite the success of perturbative QCD in describing the 
Drell-Yan cross 
sections, it remains a challenge to understand the angular 
distributions of the Drell-Yan process. Assuming dominance of the 
single-photon process, a general expression for the Drell-Yan
angular distribution is~\cite{lam78}
\begin{equation}
\frac {d\sigma} {d\Omega} \propto 1+\lambda \cos^2\theta +\mu \sin2\theta 
\cos \phi + \frac {\nu}{2} \sin^2\theta \cos 2\phi,
\label{eq:eq1}
\end{equation}
\noindent where $\theta$ and $\phi$ denote the polar and azimuthal angle,
respectively, of the $l^+$ in the dilepton rest frame. In the ``naive" 
Drell-Yan model, where the transverse momentum of the quark is ignored
and no gluon emission is considered, $\lambda =1$ and $\mu = \nu =0$ are
obtained. QCD effects~\cite{chiappetta86} and 
non-zero intrinsic transverse momentum of the quarks~\cite{cleymans81} 
can both lead to $\lambda \ne 1$ and $\mu, \nu \ne 0$. However, 
$\lambda$ and $\nu$ should still
satisfy the relation
$1-\lambda = 2 \nu$~\cite{lam78}. This so-called Lam-Tung 
relation, obtained as 
a consequence of the spin-1/2 nature of the quarks, is analogous 
to the Callan-Gross relation~\cite{callan69} 
in Deep-Inelastic Scattering. While QCD effects can significantly
modify the Callan-Gross relation, the Lam-Tung relation 
is predicted to be largely unaffected by QCD corrections~\cite{lam80}.

The first measurement of the Drell-Yan angular distribution was 
performed by the NA10 Collaboration for $\pi^- + W$ at 140, 194, and
286 GeV/c, with the highest statistics at 194 
GeV/c~\cite{falciano86}. 
The $\cos 2 \phi$ angular dependences showed a sizable $\nu$,
increasing with dimuon transverse momentum ($p_T$) and reaching a value 
of $\approx 0.3$ at $p_T = 2.5$
GeV/c (see Fig. 1). The observed behavior of $\nu$ could not
be described by perturbative QCD calculations which predict much smaller 
values of $\nu$~\cite{chiappetta86}. The Fermilab E615 Collaboration
subsequently performed a measurement of $\pi^- + W$ Drell-Yan production 
at 252 GeV/c with broad coverage in the
decay angle $\theta$~\cite{conway89}. 
The E615 results showed that
$\lambda$ deviates from 1 at large values of $x_\pi$ (the Bjorken-$x$
of the incident pions), and both $\mu$ and 
$\nu$ have large non-zero values. Furthermore, the E615 data showed that 
the Lam-Tung relation, $2\nu = 1 - \lambda$, is clearly violated. (See Fig. 1.)
\begin{figure}[tb]
\includegraphics*[width=\linewidth]{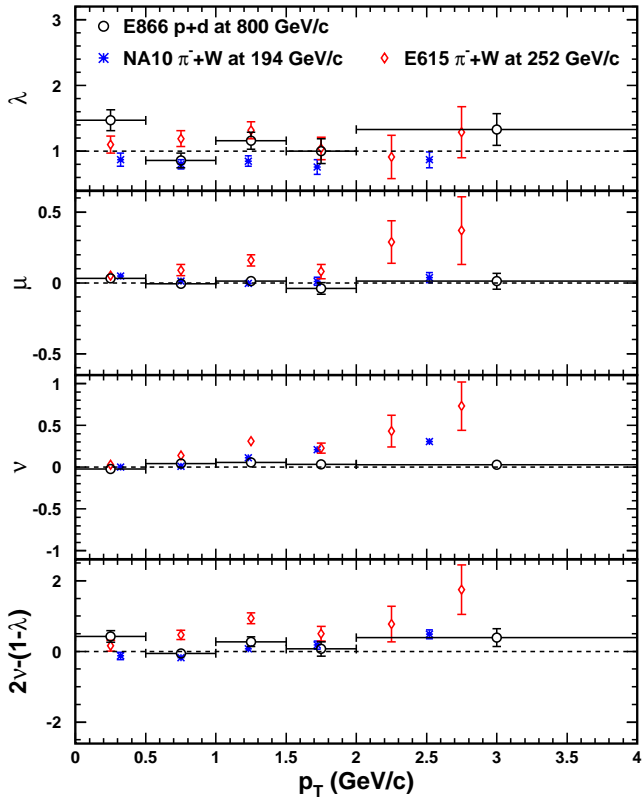}
\caption{ Parameters $\lambda, \mu, \nu$ and $2\nu - (1-\lambda)$
vs.\ $p_T$ in the Collins-Soper frame. Solid circles are for
E866 $p+d$ at 800 GeV/c, crosses are for NA10 $\pi^- + W$ at 194 GeV/c, and
diamonds are E615 $\pi^- + W$ at 252 GeV/c. The error bars 
include the statistical uncertainties only.}
\label{pdfig1}
\end{figure}

The NA10 and E615 results on the Drell-Yan angular distributions strongly
suggest that new effects beyond conventional perturbative QCD are present.
Several attempts have been made to interprete these data. Brandenburg,
Nachtmann and Mirke suggested that a factorization-breaking 
QCD vacuum may lead to a
correlation between the transverse spin of the antiquark in the pion and
that of the quark in the nucleon~\cite{brandenburg93}. 
This would result in a non-zero
$\cos 2\phi$ angular dependence consistent with the data. As
pointed out by Boer {\em et al.}, a possible 
source for a factorization-breaking
QCD vacuum is helicity flip in the instanton model~\cite{boer05}. 
Several authors 
have also considered higher-twist effects from quark-antiquark binding
in pions~\cite{brandenburg94,eskola94}, motivated by earlier work of
Berger and Brodsky~\cite{berger79}. This model
predicts behavior of $\mu$ and $\nu$ in qualitative agreement with the
data. However, the model is strictly applicable
only in the $x_\pi \to 1$ region while the NA10 and E615 data exhibit 
nonperturbative effects over a much broader kinematic region.

More recently, Boer pointed out~\cite{boer99} 
that the $\cos 2 \phi$ angular dependences 
observed in NA10 and E615 could be due to the $k_T$-dependent
parton distribution function $h_1^\perp$. This so-called Boer-Mulders
function~\cite{boer98} is an example of a novel type of $k_T$-dependent
parton distribution function, and it
characterizes the correlation of a quark's transverse spin and
its transverse momentum, $k_T$, in an unpolarized nucleon. It has an 
interesting property of being a time-reversal odd object and owes its 
existence to the presence of initial/final 
state interactions~\cite{brodsky02}. The Boer-Mulders
function is the analog of the Collins fragmentation 
function~\cite{collins93}, which
describes the correlation between the transverse spin of a quark and the
transverse momentum of the particle into which it hadronizes. 
Model calculations
for the nucleon (pion) Boer-Mulders functions have been carried 
out~\cite{gamberg03,boer03,bacchetta04,lu05}
in the framework of quark-diquark (quark-spectator-antiquark) 
model, and can
successfully describe the $\nu$ behavior observed in NA10~\cite{lu05}.

To shed additional light on the origins of the NA10 and E615 Drell-Yan
angular distributions, we have analyzed $p+d$ Drell-Yan angular
distribution data at 800 GeV/c from Fermilab E866. There are several
physics motivations for this study. First, there has been no report on
the azimuthal angular distributions for proton-induced Drell-Yan -- all
measurements so far have been for polar angular 
distributions~\cite{pat99,brown}. 
Second, proton-induced Drell-Yan data provide a stringent test of 
theoretical models. For example, the $\cos 2\phi$ dependence is expected
to be much reduced in proton-induced Drell-Yan if the underlying mechanism
involves the Boer-Mulders functions. This is due to the expectation that
the Boer-Mulders functions are small for the sea-quarks. However, if the
QCD vacuum effect~\cite{brandenburg93} is the origin of 
the $\cos 2 \phi$ angular dependence, then the azimuthal behavior of 
proton-induced Drell-Yan should be similar to that of pion-induced
Drell-Yan. Third, the validity of the Lam-Tung relation has never been
tested for proton-induced Drell-Yan, and the present 
study provides a first test. 

The Fermilab E866 experiment was performed using the upgraded Meson-East 
magnetic pair spectrometer. Details of the experimental setup have been 
described elsewhere~\cite{hawker}. An 800 GeV/c primary proton 
beam with up to $2 \times 10^{12}$ protons per beam spill 
was incident upon one of three identical 50.8 cm long cylindrical 
stainless steel target flasks containing either liquid hydrogen, 
liquid deuterium or vacuum. A copper beam dump located inside the 
second dipole
magnet (SM12) absorbed protons that passed through the target. Downstream
of the beam dump was an absorber wall that 
completely filled the aperture of the magnet.
This absorber wall removed hadrons produced in the target and the beam dump.

\begin{table}[tbp]
\caption {Mean values of the $\lambda, \mu, \nu$ parameters and the quantity
$2\nu -(1-\lambda)$ for three Drell-Yan measurements. The $p_T$ dependence
of these quantities is shown in Fig. 1.}
\begin{center}
\begin{tabular}{|c|c|c|c|}
\hline
\hline
 & $p+d$ & $\pi^- + W$ & $\pi^- +W$ \\
 & 800 GeV/c & 194 GeV/c & 252 GeV/c \\
 & (E866) & (NA10) & (E615) \\
\hline
$\langle \lambda \rangle$ & $1.07 \pm 0.07$ & $0.83 \pm 0.04$ & 
$1.17 \pm 0.06$ \\
\hline
$\langle \mu \rangle$ & $0.003 \pm 0.013$ & $0.008 \pm 0.010$ & 
$0.09 \pm 0.02$ \\
\hline
$\langle \nu \rangle$ & $0.027 \pm 0.010$ & $0.091 \pm 0.009$ & 
$0.169 \pm 0.019$ \\
\hline
$\langle 2 \nu - (1-\lambda) \rangle$ & $0.12 \pm 0.07$ & $0.01 \pm 0.04$ &
$0.51 \pm 0.07$ \\
\hline
\hline
\end{tabular}
\end{center}
\end{table}

Several settings of the
currents in the three dipole magnets (SM0, SM12, SM3)
were used in order to optimize acceptance for different dimuon mass regions.
Data collected with the ``low mass" and ``high mass" 
settings~\cite{hawker} on
liquid deuterium and empty targets were used in this analysis.
The detector system consisted of
four tracking stations and a momentum analyzing magnet (SM3). 
Tracks reconstructed by the drift chambers were extrapolated to the target
using the momentum determined from the bend angle in SM3.
The target position was used to refine the parameters of each muon track. 

From the momenta of the $\mu^+$ and $\mu^-$, kinematic variables of
the dimuons ($x_F, m_{\mu\mu}, p_T$) were readily reconstructed.
The muon angles $\theta$ and $\phi$ in the Collins-Soper
frame~\cite{collins77} were also calculated. To remove the 
quarkonium background, 
only events with $4.5 <m_{\mu\mu}<
9$ GeV/c$^2$ or $m_{\mu\mu} > 10.7$ GeV/c$^2$ were analyzed. 
A total of 118,000
$p+d$ Drell-Yan events covering the decay angular range $-0.5 < \cos\theta
<0.5$ and $-\pi < \phi < \pi$ remain. Detailed Monte-Carlo simulations
for the experiment using the MRST98 parton 
distribution functions~\cite{mrst} for
NLO Drell-Yan cross sections have shown good agreements with the data for 
a variety of measured quantities.

Figure 1 shows the angular distribution parameters $\lambda, \mu,$ and
$\nu$ vs.\ $p_T$. To extract these parameters, the Drell-Yan data were
grouped into 5 bins in $\cos\theta$ and 8 bins in $\phi$ for each $p_T$
bin. A least-squares fit
to the data using Eq. 1 to describe the angular distribution was
performed. Only statistical errors are shown
in Fig. 1. The primary contributions to the systematic errors are the 
uncertainties of the incident beam angles on target. The analysis 
has been performed 
allowing the beam angles to vary within their ranges of uncertainty. 
From this study, we found that the
systematic errors are comparable to the statistical
errors for each individual $p_T$ bin. However, the $p_T$ averaged
values $\langle\lambda\rangle, \langle\mu\rangle,$ 
and $\langle\nu\rangle$, are dominatd by the 
statistical errors.

For comparison with the $p+d$ Drell-Yan data, the NA10 $\pi^- + W$ data
at 194 GeV/c and the E615 $\pi^- + W$ data at 252 GeV/c are also shown
in Fig. 1. To test the validity of the Lam-Tung relation, also shown
in Fig. 1 is the quantity, $2\nu - (1-\lambda)$, for all three
experiments. For $p+d$ at 800 GeV/c, Fig. 1 shows that $\lambda$ is
consistent with 1, in agreement with previous studies~\cite{pat99,brown},
while $\mu$ and $\nu$ deviate only slightly from zero.
This is in contrast to the pion-induced Drell-Yan
results, in which much larger values of
$\nu$ are found. Table I lists the mean values of $\lambda, \mu, \nu$
and $2\nu - (1 - \lambda)$ for these three experiments. Again, the 
qualitatively
different behavior of the azimuthal angular distributions for $p+d$ versus
$\pi^- + W$ is evident. It is also interesting to note that while E615 clearly
establishes the violation of the Lam-Tung relation, the NA10 and the 
$p+d$ data are largely consistent with the Lam-Tung relation.

\begin{figure}[tb]
\includegraphics*[width=\linewidth]{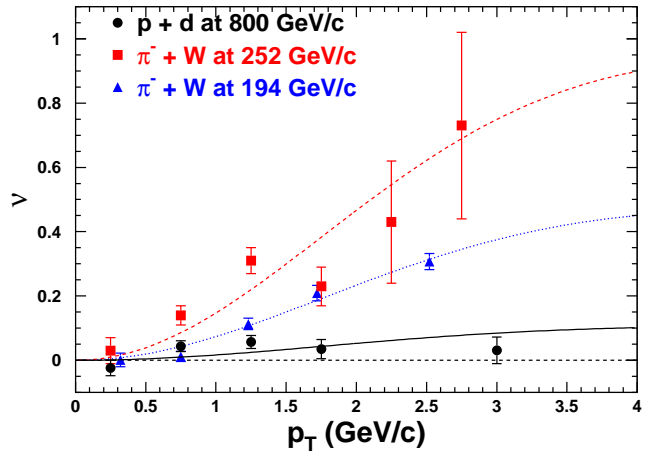}
\caption{Parameter $\nu$ vs.\ $p_T$ in the Collins-Soper
frame for three Drell-Yan 
measurements. Fits to the data using Eq. 3 and $M_C=2.4$ GeV/c$^2$ are
also shown.}
\label{pdfig2}
\end{figure}

In an attempt to extract information on the magnitude of the $h_1^\perp$
function from the NA10 data, Boer~\cite{boer99} assumed that 
$h_1^\perp$ is proportional
to the spin-averaged parton distribution function $f_1$: 
\begin{eqnarray}
h_1^\perp (x, k_T^2) = C_H \frac {\alpha_T}{\pi} \frac {M_C M_H} 
{k_T^2 + M_C^2} e^{-\alpha_T k_T^2} f_1(x), 
\label{eq:eq2}
\end{eqnarray}
\noindent where $k_T$ is the quark transverse momentum, $M_H$ is the mass 
of the hadron $H$ (pion or nucleon), and $M_C$ and $C_H$ are constant fitting 
parameters. A Gaussian transverse momentum dependence of $e^{-\alpha_T k_T^2}$
with $\alpha_T = 1$ (GeV/c)$^{-2}$ was assumed. The $\cos 2 \phi$ 
dependence then results from the convolution of 
the pion $h_1^\perp/f_1$ term with
the nucleon $h_1^\perp/f_1$ term, and the
parameter $\nu$ is given as
\begin{eqnarray}
\nu = 16 \kappa_1 \frac {p_T^2 M_C^2} {(p_T^2 + 4 M_C^2)^2},
\label{eq:eq3}
\end{eqnarray}
\noindent where $\kappa_1 = C_{H_1} C_{H_2}/2$, and $H_1$, $H_2$ denote the two
interacting hadrons. As shown in Fig. 2, a good description
of the NA10 data is obtained with $\kappa_1 = 0.47 \pm 0.14$ 
and $M_C = 2.4 \pm 0.5$ GeV/c$^2$. A fit to the E615 $\nu$ data at 252 GeV/c
using $M_C=2.4$ GeV/c$^2$, also shown in Fig. 2, gives
$\kappa_1 = 0.93 \pm 0.10$.
These large values of $\kappa_1$ suggest sizable $h_1^\perp$ functions
for the valence antiquarks in the pion and for the 
valence quarks in the nucleon.

A fit to the E866 $p+d$ data using Eq.~3 yields $\kappa_1 = 0.11 \pm 0.04$ 
for $M_C = 2.4$ GeV/c$^2$, as shown in Fig. 2. As noted earlier,
proton-induced Drell-Yan involves a valence quark annihilating with a 
sea quark. A comparison of the 
values of $\kappa_1$ from proton-induced Drell-Yan with those from
pion-induced Drell-Yan suggests that the ratio $h_1^\perp / f_1$ for 
the nucleon sea quarks is substantially below that for valence quarks.
More specifically, the 
value of $C_H$ for the sea is approximately a factor 4--8 smaller 
than that for valence quarks.

\begin{figure}[tb]
\includegraphics*[width=\linewidth]{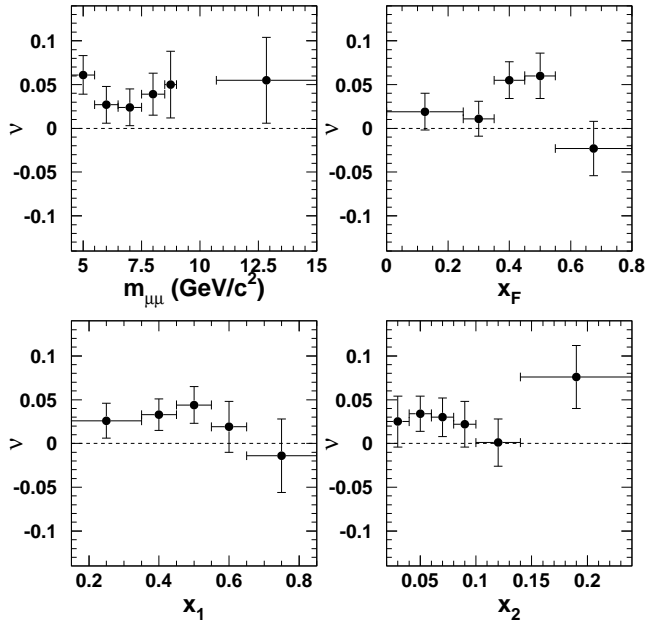}
\caption{Parameter $\nu$ vs.\ $m_{\mu\mu}$, $x_F$, $x_1$, 
and $x_2$ in the 
Collins-Soper frame for $p+d$ at 800 GeV/c. The error bars correspond
to the statistical uncertainties only.}
\label{pdfig3}
\end{figure}

The Drell-Yan angular distributions have also been analyzed for other 
kinematic variables. Figure 3 shows the values of $\nu$ for $p+d$ vs.\
$m_{\mu\mu}, x_F, x_1,$ and $x_2$, where $x_1$ and $x_2$ are the 
Bjorken-$x$ for the beam and target partons, respectively. Again, for 
each bin the data were divided into 5 bins in $\cos\theta$ and 8 bins in
$\phi$ in order to extract the angular distribution parameters. Figure 3 shows
no significant dependence on these kinematic variables.

In summary, we report a measurement of the angular distributions of
Drell-Yan dimuons for $p+d$ at 800 GeV/c. The pronounced $\cos 2 \phi$
azimuthal angular dependence observed previously in pion-induced Drell-Yan
is not observed in the $p+d$ reaction. The Lam-Tung relation, found to be
strongly violated in the E615 pion-induced Drell-Yan, remains largely 
valid for 
$p+d$ Drell-Yan. These results put constraints on theoretical models that 
predict large $\cos 2 \phi$ dependence originating from QCD vacuum effects.
They also suggest that the Boer-Mulders functions 
$h_1^\perp$ for sea quarks are significantly smaller than those for 
valence quarks.

This work was supported in part by the U.S. Department of Energy and 
the National Science Foundation.

\end{document}